\documentclass[epj]{svjour}
\usepackage{graphics}
\usepackage{epsfig}
\input epsf
\def\sfn{scale free networks }

\begin{document}

%\centerline{\bf \huge{Bounded confidence and Social networks }}

% \bigskip
%\centerline{\bf G{\'e}rard~Weisbuch}

%% \bigskip
%\centerline{Laboratoire de Physique Statistique$^4$}

%% \bigskip
%\centerline{de l'Ecole Normale Sup{\'e}rieure, }

%%\bigskip
%%\centerline  {24 rue Lhomond, F 75231 Paris Cedex 5, France. }

%%\noindent
%%\bigskip
%%{\small $^4$Laboratoire associ{\'e} au CNRS (URA 1306), {\`a} l'ENS
%%et aux Universit{\'e}s Paris 6/7 }

%%\bigskip
%%\centerline{e-mail: weisbuch@lps.ens.fr}

\title{Bounded confidence and Social networks}

\author{G{\'e}rard~Weisbuch}

\institute{Laboratoire de Physique Statistique\footnote{Laboratoire associ{\'e} au CNRS (URA 1306), {\`a} l'ENS\\
et aux Universit{\'e}s Paris 6/7} \\
de l'Ecole Normale Sup{\'e}rieure,\\
24 rue Lhomond, F 75231 Paris Cedex 5, France.\\
\email {weisbuch@lps.ens.fr}
}

%%\email {weisbuch@lps.ens.fr}
\date{Received: date / Revised version: date}

\abstract{In the so-called
bounded confidence model 
proposed by Deffuant et al, agents
can influence each other's  
opinion provided that opinions are already 
sufficiently close enough.
  We here discuss the influence of
possible social networks topologies 
on the dynamics of this model.
}


\maketitle

\section{Introduction}
\label{sec:intro}

  Many models about opinion dynamics, 
\cite{Follmer}, 
\cite{Galam}, \cite{Arthur},
are based on binary opinions
which social actors update as a result of social influence,
often according to some version of a majority rule.
  Binary opinion dynamics
have been well studied, such as the herd behaviour described 
 by economists (
\cite{Follmer}, \cite{Arthur}, \cite{Orlean}).
When binary interactions can occur about any pair of agents
randomly chosen,
the attractors of the dynamics display uniformity of
opinions, either 0 or 1.
 Clusters of opposite opinions appear when the 
dynamics occur on a social network with exchanges restricted
to connected agents. These patterns remind of magnetic domains in
Ising ferromagnets.

 The spreading of epidemics on \sfn \cite{BA} is also an instance of
a binary state dynamics \cite{Vespignani}.

 One issue
of interest concerns the importance of the binary assumption:
what would happen if opinion were a continuous variable such as 
the worthiness of a choice (a utility in economics), or some belief
about the adjustment of a control parameter? 
 These situations are encountered in economic and
social science:
\begin{itemize}
\item In the case of technological changes economic agents
have to compare the utilities of a new technology with respect
to the old one, and e.g. surveys concerning the adoption
of environment friendly practicies following the 1992
new agricultural policies \cite{images}
showed that agents have uncertainties about
the evaluation of the profits when they adopt the 
new technique  and thus partially rest on evaluations 
made by their ``neighbours''. 
\item Some social norms such as how to share the 
profit of the crop among landlords and tenants \cite{py} do display
the kind of clustering that we will further describe.
\end{itemize}

In the
bounded confidence model of continuous opinion dynamics
proposed by Deffuant etal \cite{Deffuant}, agents
can influence each other's  
opinion provided that opinions are already 
sufficiently close enough. A tolerance
threshold $d$ is defined, such that
agents with difference in opinion
larger than the threshold can't interact. 
Several variants of the model have been proposed 
in \cite{Deffuant} \cite{Hegselmann}.
In these models, the only restriction for interaction 
is the threshold condition and interactions among  
any pair of agents can occur.
 The attractor of the dynamics are clusters
which number increases by steps when the tolerance threshold 
is decreased.

 The dynamics which we will describe here can be compared 
to the cultural diffusion model introduced
by Axelrod: agents culture is represented by 
strings of integer in these models \cite{axel}.

The purpose of this paper
is to check the role of specific interaction structures
on the result of the dynamics. We will investigate
a bounded confidence interaction process on  
scale free networks and compare the obtained dynamics to what was
already observed when all interactions are possible
and when they occur on square lattices among nearest neighbours.

The paper is organised as follows:
\begin{itemize}
\item 
We first expose the simple case of complete mixing 
among agents.
\item 
We then check the genericity of the results obtained
for the simplest model to other topologies,
mostly scale free networks.
\end{itemize}

  We are mainly interested in:
  \begin{itemize}
  \item the clustering process,
  \item the possible existence of regime transitions
according to the value of the threshold of influence $d$  
   \item the relative importance of the clustering process
with respect to the whole population. Do all 
or at least most agents participate into 
this process? 
\end{itemize}

\section{The basic case: Complete Mixing}
  
  Let us consider a population of $N$ agents $i$
with continuous opinion $x_i$. We start from an initial distribution 
of opinions, most often taken uniform on [0,1] in the computer simulations.
 At each time step any two 
randomly chosen agents
meet:
 they re-adjust their opinion when their difference in 
 opinion is smaller in magnitude than a threshold $d$.
Suppose that the two agents have opinion $x$ and $x'$.

$Iff$ $|x-x'|<d$ opinions are adjusted according to:
\begin{eqnarray}
  x &=& x + \mu \cdot (x'-x) \\
  x' &=& x' + \mu \cdot (x-x') 
\label{eq1}
\end{eqnarray}

 where $\mu$ is a convergence rate
whose values may range from 0 to 0.5.

In the basic model \cite{Deffuant}, the threshold $d$ is taken as
 constant in time and across the whole population. 
Note that we here apply a complete mixing hypothesis
plus a random serial iteration mode\footnote{The "consensus" 
literature \cite{Hegselmann} most often uses parallel iteration mode when they
suppose that agents average at each time step the opinions
of their neighbourhood. Their implicit rationale
 for parallel iteration is that they model successive
 meetings among experts. 
}.

  For finite thresholds,
computer simulations show that the distribution of opinions
evolves at large times towards clusters of homogeneous opinions.
The number of clusters varies as the integer part of $1/2d$:
this is to be further referred to as the "1/2d rule"
(see figure 1\footnote{Notice the continuous transitions
in the average number of clusters when $d$ varies. 
Because of the randomness of the initial distribution and pair
sampling, any prediction on the outcome of dynamics
such as the 1/2d rule only becomes true with a probability close
to one in the limit of large $N$.}).
\begin{figure}[!ht] 
\centerline{\epsfxsize=90mm\epsfbox{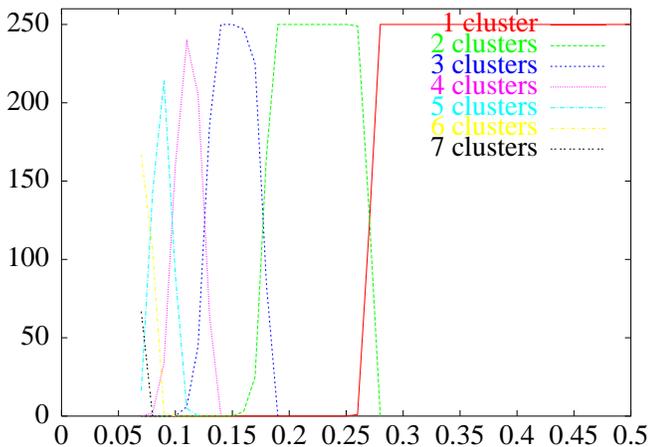}}
\caption{\label{fig:phases} Statistics of the number of opinion clusters
as a function of $d$ on the x axis for 250 samples ($\mu=0.5, \quad
N=1000$).} 
\end{figure}

\section{The scale free network topology and 
opinion updating process}

  We use a standard method, see e.g. Stauffer
and Meyer-Ortmanns \cite{HMO}:
 
 Starting from a fully connected network of
3 nodes, we add iteratively nodes (in general up to
900 nodes) and connect 
them to previously created nodes in proportion
to their degree. We have chosen to draw two 
symmetrical connections per new added node in order to achieve
the same average connection degree (4) as in the 30x30 square
lattice taken as reference. But obviously the obtained networks
are scale free as shown by Barabasi and Albert\cite{BA}.  

  In fact \sfn \cite{BA} display a lot of heterogeneity in nodes connectivity. 
In the context of opinion dynamics,
well connected nodes might be supposed more influential,
but not necessarily more easily influenced. At least this is 
the hypothesis that we choose here.
 We have then assumed asymmetric updating:
a random node is first chosen, and then one of
its neighbours. But only the first node in the pair
might update his position according to equ.1, not both. As a result,
well connected nodes are influenced as often as 
 others, but they influence others in proportion
to their connectivity. This particular choice 
of updating is intermediate between what
Stauffer and Meyer-Ortmanns \cite{HMO} call directed and
undirected versions.

\section{Clustering and transitions}
   
  A simple way to check clustering, and especially
on average, for any topology is the dispersion index $y$ proposed by
Derrida and Flyberg \cite{Derrida}.
  $y$ is the relative value of the ratio of the sum
of the squared cluster sizes $s_i^2$
 to the squared number of agents.

\begin{equation}
  \label{eq:xx}
  y = \frac{\sum_{i=1}^n s_i^2}{(\sum_{i=1}^n s_i)^2}
\end{equation}

 For $m$ clusters of equal size, one would have $y=1/m$.
 The smaller $y$,
the more important is the dispersion in opinions.

\begin{figure}[hbt]
%%\begin{center}
%%\rotatebox{270}{\scalebox{0.5}{\includegraphics{deffuantparis6.ps}}}
%%\end{center}
\centerline{\epsfxsize=90mm\epsfbox{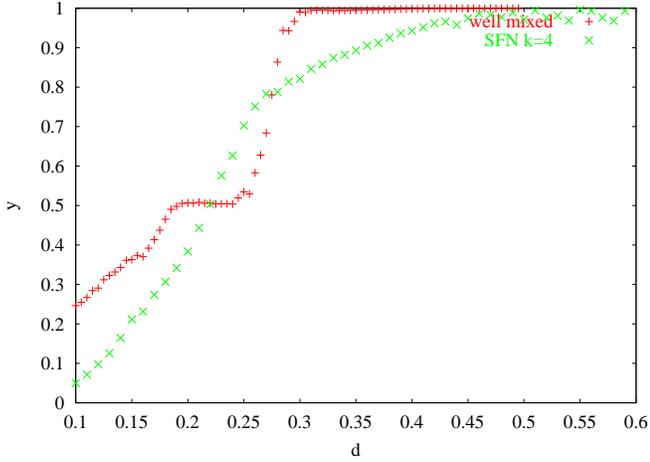}}
\caption{ \label{figaver} Dispersion index $y$ as a function of the
tolerance threshold $d$ for well mixed systems (red '+')
\sfn (green 'x') with 900 nodes. Each data 
point is the result of an average over 100 simulations.}
\end{figure}

   When averaging over network topology and initial conditions
the step structure (fig. \ref{figaver}) observed in the case of full
mixing seems to be completely blurred. For  \sfn
one observes
a continuous increase of the Derrida Flyberg parameter
as a function of the tolerance threshold with only
 a kink in the $d=0.25, y=0.7$
region; while two distinct steps at $y=0.5$ and $y=0.33$ 
are observed in the well mixed case,
corresponding to the occurence of 2 and 3 large
clusters respectively.

    In fact the blurring of the transition 
in \sfn is due to two effects:
\begin{itemize}
\item 
the S curve is the result of  averaging
 over many network topologies and initial conditions.
\item 
 Presence of outlying\footnote{During the iterative process of opinion
 exchange, nodes with few connections have less chances to interact
 with a neighbour which opinion is close enough 
from their own opinion to actually interact. Many of them are not
 affected by the convergence process and remain outside the
 distribution of clustered opinions. We call them
outlying nodes.} nodes \cite{Neau} in \sfn, which remain out
of the clustering process, decrease $y$, especially at low tolerance
values.
\end{itemize}

\begin{figure}[htbp]
\centerline{\epsfxsize=90mm\epsfbox{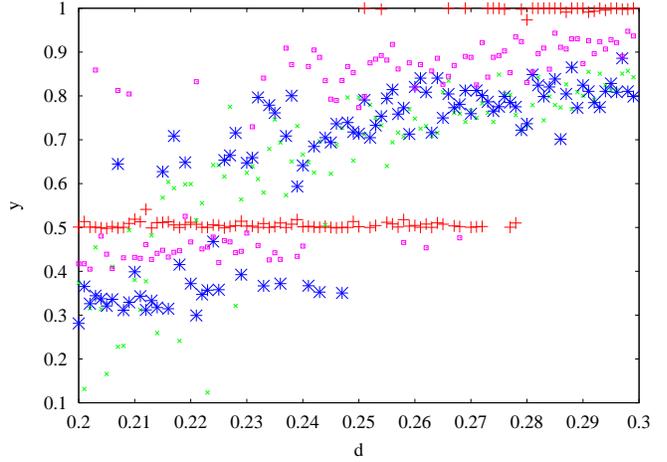}}
\caption{\label{fignoaver}  Variation of the dispersion index $y$ as 
a function of the tolerance threshold $d$. 
 Big red '+' correspond to the well-mixed case,
 small green 'x' to square lattice ,  big blue '*' to scale free network
with connectivity 4 and  small violet squares to scale free network
with connectivity 8}
\end{figure}

  When measurements are done on single instances of network topology 
and initial conditions, one observes $y$ values corresponding to 
either one (larger $y$ values) or two clusters (smaller $y$ values)
in the $0.2 <d <0.3$ region. The proportion of these two 
$y$ values varies with $d$, larger $y$ values being 
more often obtained with 
larger $d$ values. For the sake of comparison figure \ref{fignoaver}
displays the variations of the  dispersion index with the tolerance
threshold for three different topologies: the standard well-mixed case
where any agent might interact with any other one, the square lattice
and the scale free network with an average connectivity $k$ equal
to 4 and 8 ($k=4$ is the same as the connectivity of the square lattice).  

   One observes that in the well mixed case the $y$ values are either 0.5 or 
1, with a rather narrow ambiguous region in $d$. For scale free
networks, $y$ values are smaller,
 an indication of the existence of many outlying agents which 
opinion does not cluster 
because they are too isolated (see further).
Their distribution looks bimodal in a larger ambiguous region.
 The magnitude and dispersion of $y$ values
is similar for  scale free network
with connectivity 4 and square lattices.
Increasing the average connectivity by a factor 2
brings the scale free network results closer to those of
the well-mixed case. Connectivity at this stage seems
more important than topology. 

\begin{figure}[htbp]
\centerline{\epsfxsize=90mm\epsfbox{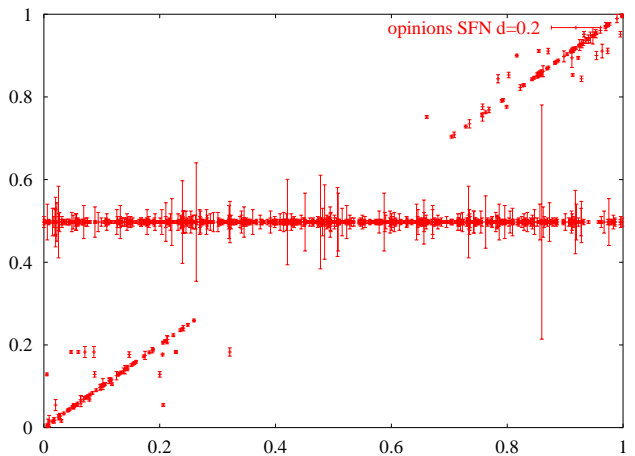}}
\caption{ \label{fig:xy} Final opinions
versus initial opinions on a scale free network with average connectivity 4
and tolerance 0.2. Vertical bars give the number of neighbours
of each node (the largest correspond to 85).}
\end{figure}

    One of the most important questions in \sfn is the role of 
the most connected nodes with respect to the less
connected ones. In the context of opinion dynamics,
we might want to figure out whether they are 
more influential, or eventually more influenced?
One answer is provided by checking how far their
opinion is changed by the clustering process.
 Figure \ref{fig:xy} is a plot of final opinions of agents as a 
function of their initial opinion. Nodes connectivity
are indicated by the size of the vertical bars.
 The importance of clustering is indicated by the density of
points on horizontal lines while outlying agents are 
located on the first bisectrix.

  Most of the well connected nodes belong to horizontal
cluster at $x_{\infty}=0.5$. They are far from the first bisectrix, which imply
that they have been influenced during the clustering process.

The first bisectrix is composed of less connected nodes,
which initial and final opinion are more than
$d=0.2$ away from the cluster. These nodes have not changed their
opinion. In scale free networks, static isolation (due to lower
connectivity) often results in being kept out of
the clustering process and remaining outlying.
The effect is systematically observed for all tolerance thresholds
less than 0.5. The outlying number explains
 why the highest values of $y$ are lower than 1
in figure 3: only
one central cluster is present, but it only contains a fraction
of the nodes.

  For the same $k$ values, 
well mixed systems display horizontal clusters in this 
[$x_0,x_\infty$] representation but very few outlying agents.
 Their occurence relates to dynamics:
when the dynamics is fast some agents remain outlying
 when they are reached for a possible updating after the
convergence process has been already well engaged, because of the
randomness of the iteration process. Agents with
initial extreme values have more chances to become outlying,
but those who actually do, depend upon the particular instance
of the random iteration.    

  Stauffer et al \cite{HMO} have done extended statistics of the total
number of different opinions after convergence 
in \sfn. Since the number of outlying agents is much bigger 
than the number of big clusters, their figures
give a very good characterisation of the number of outlying nodes.
  
\begin{figure}[htbp]
\centerline{\epsfxsize=90mm\epsfbox{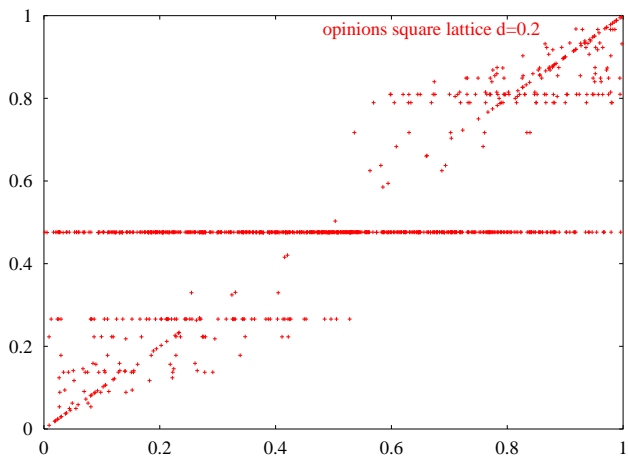}}
\caption{ \label{figxycar} Final opinions
versus initial opinions on a 30x30 square lattice
with tolerance 0.2. }
\end{figure}

For the sake of comparison we give the equivalent display for
square lattices (fig. \ref{figxycar}). 
The results are pretty similar to those obtained with 
\sfn. The less populated horizontal lines correspond
to small connected clusters on the lattice.

\section{Conclusions}

  In conclusion, restricting influence by
a network topology does not drastically change
the behaviour of these models of social
influence as compared to the well mixed case.
 To summarize some of the resemblances and differences:
 
\begin{itemize}
\item  One does observe clustering effects,
and the number of observed main clusters 
does not largely differ for what is observed 
for equivalent tolerance thresholds in
 the well mixed case. Caution: we have only been discussing
clusters in terms of opinions, not in terms of 
connections across the network. For small $d$
values, clustering in opinion might structure
the network in smaller connected regions  
with clustered opinions. One can expect the
number of such non-interacting regions to be larger 
than the number of clusters (as observed on
square lattices \cite{Deffuant}).

\item Stairs of $y$, the dispersion index, do appear:
at least when measured without averaging 
on single instances of networks and initial conditions.
But $y$ values are decreased by a larger proportion of 
outlying agents and the transition regions in tolerance are larger.   

\item  Well connected nodes are influenced by
other nodes and are themselves influential. 
Most of them belong to the big
cluster(s) after the clustering process.
 
\item Larger connectivities bring \sfn dynamic
behaviour closer to well mixed systems.

\begin{acknowledgement} 
 We thank D. Stauffer for helpful discussions and 
early communication of his results \cite{HMO},
and F. Amblard for useful remarks.
 We acknowledge partial support from the FET-IST
grant of the EC IST 2001-33555 COSIN.
\end{acknowledgement}

\end{itemize}

\end {document}